\documentclass{article}
\usepackage{microtype}
\DisableLigatures{encoding = *, family = * }
\usepackage{amsmath}
\usepackage{algorithm}
\usepackage{algpseudocode}
\usepackage{algpascal}
\usepackage[hidelinks]{hyperref}
\usepackage{algorithm}
\usepackage{algpseudocode}
\usepackage{algpascal}
\usepackage{subfigure}
\usepackage{float}
\usepackage{makeidx}
\usepackage{balance}
\usepackage{graphicx}
\usepackage{lscape}
\usepackage{multirow}
\usepackage{multicol}
\usepackage{verbatim}
\usepackage{subfigure}
\usepackage{newclude}
\usepackage{array}
\usepackage{booktabs}
\usepackage{graphicx}
\usepackage{caption}
\usepackage{acronym}
\usepackage{listings}
\usepackage{booktabs}
\usepackage{colortbl}
\usepackage{xcolor}
\usepackage{microtype}

\usepackage{pgfplots}
\pgfplotsset{compat=newest}
\usetikzlibrary{plotmarks}
\usetikzlibrary{shapes,arrows}

\tikzstyle{decision} = [diamond, draw,
    text width=5.5em, text badly centered, node distance=3cm, inner sep=0pt]
\tikzstyle{block} = [rectangle, draw,
    text width=7.5em, text centered, rounded corners, minimum height=4em, node distance=2cm, fill=gray!10]
\tikzstyle{rect} = [rectangle, draw,
    text width=10.5em, text centered,  minimum height=4em, node distance=2cm, fill=gray!40]
\tikzstyle{line} = [draw, -latex']
\tikzstyle{cloud} = [draw, ellipse, node distance=4cm,text width=4.5em,
    minimum height=2em]

\newlength\figureheight
\newlength\figurewidth
\acrodef{HMM}{Hidden Markov Models}
\acrodef{FHMM}{Factorial Hidden Markov Model}
\acrodef{NILM}{Non-Intrusive Load Monitoring}
\acrodef{PF}{Particle Filtering}
\acrodef{ETC}{effort-to-compress}
\acrodef{NSRPS}{non-sequential recursive pair substitution}
\acrodef{PC}{power similarity complexity}
\acrodef{C}{combined complexity}
\acrodef{ACC}{accuracy}
\acrodef{PDF}{probability density function}
\acrodef{RMSE}{root mean squared error}

\usepackage{siunitx}
\usepackage{tikz}
\usepackage{pgfplots}
\pgfplotsset{width=7cm,compat=newest}

\begin{document}


\title{Complexity of Power Draws for Load Disaggregation}

\author{
Dominik~Egarter, Manfred P\"ochacker and~Wilfried~Elmenreich\\
Institute of Networked and Embedded Systems\\
 Alpen-Adria-Universit\"at Klagenfurt, Austria\\
 \{\emph{name.surname}\}@aau.at
 } 

\maketitle              


\maketitle
 \begin{abstract}
\ac{NILM} is a technology offering methods to identify appliances in homes based on their consumption characteristics and the total household demand.
Recently, many different novel \ac{NILM} approaches were introduced, tested on real-world data and evaluated with a common evaluation metric.
However, the fair comparison between different \ac{NILM} approaches even with the usage of the same evaluation metric is nearly impossible due to incomplete or missing problem definitions.
Each \ac{NILM} approach typically is evaluated under different test scenarios. Test results are thus influenced by the considered appliances, the number of used appliances, the device type representing the appliance and the pre-processing stages denoising the consumption data.
This paper introduces a novel complexity measure of aggregated consumption data providing an assessment of the problem complexity affected by the used appliances, the appliance characteristics and the appliance usage over time.
We test our load disaggregation complexity on different real-world datasets and with a state-of-the-art \ac{NILM} approach.
The introduced disaggregation complexity measure is able to classify the disaggregation problem based on the used appliance set and the considered measurement noise.
\end{abstract}


\section{Introduction} \label{sec:introduction}

The power draw of a household is composed by aggregated power profiles of each household appliance.
By knowing the household power draw as well as the appliance characteristics, such as power consumption, it is possible to disaggregate the household power draw into its used appliance components.
\ac{NILM}, also known as load disaggregation or non-intrusive appliance load monitoring\footnote{The terms \ac{NILM} and load disaggregation are used in the same context and are replaceable throughout this paper}, was firstly introduced by Hart \cite{Hart1992} in 1992 solving the problem to disaggregate load profiles provided by different techniques and algorithms.
The up to now proposed \ac{NILM} algorithms cannot solve the problem of \ac{NILM} in all its aspects.
To be able to improve the state-of-the-art of load disaggregation approaches, it is necessary to compare different algorithms in a fair way.
Unfortunately, a fair comparison between different algorithms is not possible due to the fact that recent approaches are highly dependent on different conditions and features such as:
\begin{itemize}
\item sampling frequency of the household power draw,
\item number of observed appliances,
\item appliance types (e.g., on/off appliances, multi-state appliances),
\item the filtering approach applied to the household power draw (the power draw feed into the \ac{NILM} algorithm is usually filtered and preprocessed before evaluations) and
\item set of used appliance features (e.g., steady state electrical characteristics, transient behavior, etc.).
\end{itemize}
Therefore, a possible comparison between algorithms is possible as for example how many feature are used or on which sampling frequency is the algorithm able to work.
But an algorithm comparison lacks of the ability to compare and to evaluate the results of the load disaggregator even if the same dataset was used.
There is the need of a common quantitative merit for \ac{NILM} which is algorithm independent and considers data assumption as well as data pre-processing.
Thus, the problem itself has to be made comparable which is created by the used appliances in a house, their appliance characteristics and their usage over time.
A possibility to make the load disaggregation problem comparable is to describe the complexity of the problem in which the problem can be seen as a simple time series.
To describe the complexity of time series different complexity measure were proposed such as entropy-based complexity measures \cite{Pincus1991,Richman2000,Rosso2003}, used for different applications such as DNA \cite{Monge2014,Costa2005} sequences or EEG \cite{Rezek1998,Jäntti2004,Gao2012} signals.
The problem of load disaggregation is hard due to the high variety of different appliances, their different ways to consume energy and their high time-variant behavior introduced by the appliance user.
It is therefore necessary to involve appliances and their characteristics as well as the time dependent behavior into the evaluation of a possible complexity measure.

In this paper, we propose an approach to make the disaggregation problem of aggregated power demands comparable by introducing two novel load disaggregation complexity measures.
To the best of our knowledge, this is the first approach summarizing  the disaggregation problem as a complexity value created by statistical characteristics of the appliance set and the time series behavior.
A similar approach concentrating the fundamental limits of \ac{NILM} was introduced in \cite{Dong2014}.
In this paper the authors derive an upper bound on the probability to distinguish scenarios for an arbitrary \ac{NILM} algorithm to guarantee on when \ac{NILM} is impossible without using privacy ensuring approaches as presented in \cite{prokop14}.
The work in \cite{Dong2014} differs from our approach as we try to make the problem of superimposed loads with the used appliance characteristics comparable between different used \ac{NILM} algorithms.
The two proposed disaggregation complexity merits are evaluated on real-world data and compared to the disaggregation result of a state-of-the-art \ac{NILM} algorithm.

The remainder of this paper is organized as follows: In Section \ref{sec:Complexity} the disaggregation complexity of aggregated power draws and factors influencing this complexity are identified.
With this knowledge an appliance set complexity and time series complexity is defined in Section \ref{sec:appCompl}.
In Section \ref{sec:evaluationSettings} the used appliance datasets, the way to extract possible power states out of measurement data and a possible load disaggregation approach used for evaluations are defined, followed by Section \ref{sec:caseStudy} presenting three case studies reviewing the complexity measures according to their suitability and meaningfulness to describe the load disaggregation problem. The approach and the results are discussed in Section \ref{sec:discussion}. Section \ref{sec:conclusion} concludes the paper.

\section{Complexity of the Power Draw makes Hardness for Disaggregation} \label{sec:Complexity}
%

\begin{figure}[h!]
\centering
\includegraphics[width=.7\columnwidth]{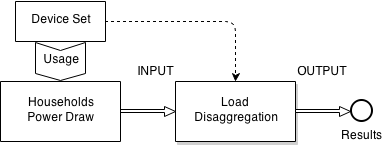}
\caption{A performance metric for load disaggregation should only compare its input with its output. The complexity of the input, which is the power draw, must be assessed separately. }
\label{fig:NILM}
\end{figure}

\begin{figure*}[th]
\centering
\includegraphics[width=1\columnwidth]{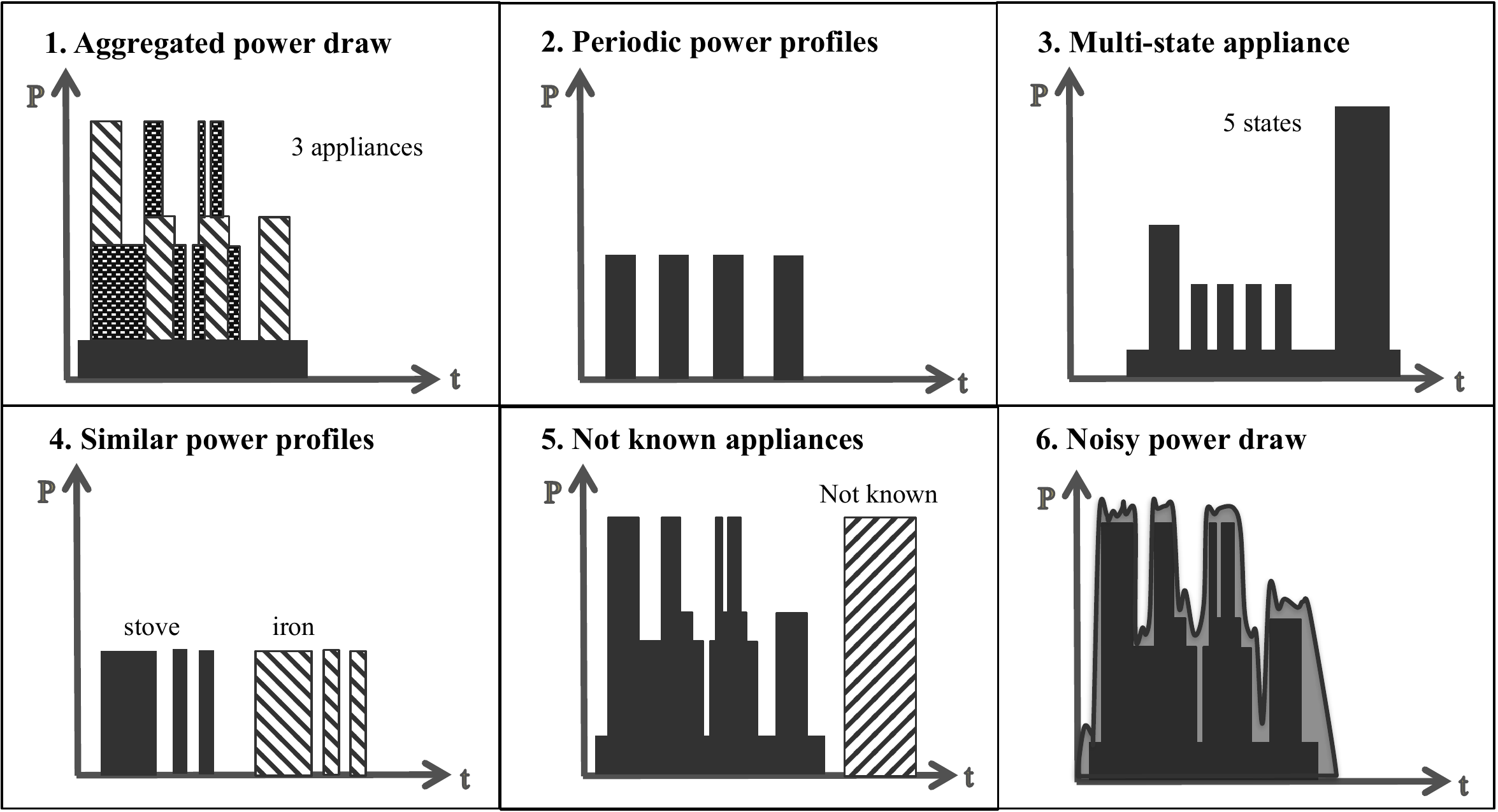}
\caption{Overview of different scenarios and characteristics of aggregate power draws which are increasing the complexity of disaggregating the household power demand.}
\label{fig:HEMSOverview}
\end{figure*}

The problem of load disaggregation is to break down the household power draw $P(t)$ to its power consumption components $p_n(t)$.
This can be formulated as the superimposition of the appliance power profiles over time as
\begin{equation}
P(t) = p_1(t) + p_2(t) + \dots + p_N(t) \text{ for } t \in \{1,T\}
\end{equation}
where $N$ represents the number of used appliances.
Each power profile $p_N$ has its own behavior to consume energy determined by the appliance power states (e.g.: on/off appliance, multi-state appliance) and the appliance usage (e.g.: fridge with periodic usage, TV with common usage times) over time.
The task of a load disaggregator is it then to find the best combination of known appliance power profiles to minimize the error between the estimated power signal and the household power draw.
The computational complexity theory can be used to describe the way and complexity to find the best solution. The theory of computational complexity is widely applied to quantify the difficulty or hardness of computational problems and state whether a (type of) problem is solvable at all and how the calculation time scales with the problem size.
In that sense load disaggregation is shown to be NP hard by Hart \cite{Hart1992}.
By combining this knowledge the NILM process is sketched in Figure \ref{fig:NILM}.
The input for the literal load disaggregation is the (households) power draw $P(t)$ that is generated by the usage of devices.
Characteristics of the single devices are known and used by to the load disaggregator.
The device characteristics can be learned, be known a-priori or entered by expert knowledge.
The device usage is the unknown part of the \ac{NILM} process that generates the power draw which can be simple or hard to disaggregate, even for the same set of devices.
Therefore, a complexity measure of load disaggregation has to be able to handle this circumstance as well as the problems stated in Figure \ref{fig:HEMSOverview}.
These problems include:
\begin{enumerate}
\item The complexity of aggregated loads is increasing with increased number of appliances due to higher probability of ambiguous power draws.
\item The higher the switching frequency (like in the case of periodic performing appliances such as a fridge), the more complex is a device set for a load disaggregation algorithm.
\item Appliances with several operation states (i.e., multi-state appliances instead of simple on/off appliances) make a device set more complex for a load disaggregation algorithm.
\item The higher the similarity between appliance features, the more complex is the problem. Similarity features are for example state power values or consumption shapes.
\item Additional noise, unknown or not considered appliances interfere with the household power draw and increases the complexity of the problem because the presence of noise typically increases the number of possible interpretations of a power draw.
\end{enumerate}
The aim is now to define a complexity measure describing the load disaggregation problem by a comparable quantity without taking the used load disaggregator into account.
The complexity measure should be independent from the used load disaggregation approach and describe the problem of aggregated power loads.
The problem is defined by the input to the load disaggregator which is the household power draw and the set of appliances.
The set of appliances is defined by the appliance consumption behavior in steady state.
Therefore, appliances are described by power consumption states as for example an on/off appliances are described by two states representing an appliance to be on or off.

As a first approach to describe the complexity of households power draws, the well-known Shannon Information could be considered.
The Shannon Information or Entropy \cite{Cover2005} is defined for either a stream of symbols or a source in general, if the symbol occurrence probabilities are known (or can be assumed).
For a specific source it means the averaged information of all possible streams.
It is developed for communication theory and is not directly part of computational complexity theory.
In case of NILM the power draw can be interpreted as a stream of symbols.
The set of possible symbols is then defined by the power values of single devices and their possible combinations, respectively.
Entropy reflects the difficulties in NILM related to the number of involved appliances.
Noise could be included in a continuous formulation of Shannon's entropy, but the problems related to very similar or equal power values for different states are not reflected.
Further there is the problem that the concept states about averages of all possibilities or so called typical sequences. 
So far it is unknown whether load profiles are typical in that sense .\newline
Another approach to describe complexity is to use the Kolmogorov complexity\cite{Cover2005}.
It is the idea to describe the complexity of a stream by the length of the shortest possible program that can generate this specific stream that is widely used in computer science.
It is a theoretical concept but there is currently no general method to estimate it.
It can be well approached in practice but it remains the uncertainty about existence of a shorter (undiscovered) solution.
In this context the device usage in the NILM process is interpreted as a program that is producing the stream.
The disaggregation algorithm would be somehow an "inverting" program.
A periodic device profile is not very complex in this sense, still many NILM approaches have difficulties in its disaggregation.
The average Kolmogorov complexity of all possible streams approaches the Shannon Information as shown by \cite{Cover2005, Gruenwald2004}.
The specification of load disaggregation problems requires a complexity measure that describes a specific power draw as the Kolmogorov complexity and is calculable such as the Shannon entropy.
Due to the fact that standard complexity theories as the Shannon entropy and the Kolmogorov complexity fails to entirely describe the load disaggregation problem, a new complexity measure has to be introduced which should fulfil the following requirements:
\begin{enumerate}
\item Describes the load disaggregation problem and should not be dependent on the load disaggregation approach.
\item Includes appliance descriptions as number of states and the similarities between appliances and states.
\item Should be applicable to time series to describe the influence of appliance usage affecting the used \ac{NILM} approach.
\item Should be easy and understandable as standard complexity theories.
\item Must not be a general complexity merit. It is a application dependent complexity measure to make load disaggregation problems comparable without considering the load disaggregator.
\end{enumerate}

\section{Novel Complexity Measure for Load Disaggregation}\label{sec:appCompl}

The proposed complexity measure should reflect the complexity of the load disaggregation problem and therefore, should make \ac{NILM} problems comparable.
In this work we follow the idea that each possible produced power value is a combination out of all possible power states of appliances.
This results in the task for the load disaggregator to find the best matching combination of power states with the measured power values.
The measured power value is influenced by noise and should be approximated as good as possible enabling the load disaggregator to decide which appliances are running.
The main idea is to relate an observed power value to all possible power state combination under the influence of measurement noise.

\subsection{Appliance Set Complexity}
One of the major factors influencing the complexity of aggregated power profiles, is the set of possible power values.
The more complex the appliance model and their operational states are, the more complex is the problem to disaggregate the power profiles.
In general, the appliance set is composed by $N$ different appliances. 
With the knowledge of the appliance set and power demands of each appliance, the first step is to compute the number of possible aggregated power values $M$.
In case of only two state devices are $2^N$ combinations, generally are
\begin{equation}
M = 2^{N_2} 3^{N_3} \dots = \prod_{Z=2}^{Z_{max}}{ Z^{N_Z}}
\label{eq:possibleStates}
\end{equation}
different power values possible. $N_2$ is the number of appliances with two states, $N_3$ with three states and so forth.
For the calculation of all possible aggregated power values $P_i$, repetitions of the same value are possible for instance if a water kettle and a coffee machine consume the same power.
Exceptions are the 0W power state (all off) and the all-on-state $P_M$ which is the highest possible power value.
The vector $P$ is the set of all possible (aggregated) power values $P_i$ for a set of appliances, where $i$ is defined as $i \in [1,M]$.

In its simplest form a NILM device observes a power value and compares it to all possible values $P_i$ given by the device set.
As long as there is one single matching power value in the set the task is solved straight forward.
The problem is harder if either are two or multiple matching values or if the value is not in the set at all.
For the disaggregation complexity measure we reason that it should contain something like a multiplicity or occupation number of the possible power values to reflect multiple occurrence. The second case does not occur in ideal NILM problems but in reality it is likely that a measured power value does not match exactly to any of the $M$ aggregated power values.
Therefore, we propose to represent the possible power values by a distribution function instead of a single value.
Following this, it is possible to estimate for a power value, which would not be in the discrete set, the probability for being caused by the respective state.
This approach covers also uncertainties caused by adjacent power values which hardly can be distinguished, e.g. through insufficient measurement accuracy in the NILM device.
A simple measure for the similarity of two distributions is the overlapping coefficient 
\begin{equation}
\mathrm{OVL}(f_1,f_2) = \int_{x} \mathrm{min} ( f_1(x),f_2(x) ) \mathrm{d}x
\end{equation}
which gives the intersection area of the two distribution curves $f_1$ and $f_2$ as stated in \cite{Inman1989}.

For a load disaggregation complexity measure $C$ we propose to estimate the similarity of one power value distribution to all the other possible aggregated power valued distributions.
The possible power values are expected between $0$ and $P_M$. By use of the overlapping coefficient the disaggregation complexity measure for the power state $P_k$ is defined as
\begin{equation}
	\begin{split}
		C_{k} = & \sum_{j=1}^M \mathrm{OVL}(f_{P_k},f_{P_j})  \\
				 =  &\sum_{j=1}^M \int_{0}^{P_M} \mathrm{min}(f_{P_k}(p),f_{P_j}(p))\mathrm{d}p \quad.
	\end{split}
\end{equation}
$C_{k}$ is the disaggregation complexity of the power value $P_k$ within the set of $M$ power state combinations.
The parameter $k$ determines the chosen reference power state combination, where $k \in [1,M]$.
In case the exact distribution of the power values are not know it is reasonable to assume a normal-distributed \ac{PDF} $ \mathcal{N} (\mu,\sigma)$. The mean value $\mu = P_k$ represents the observed power value and a variance $\sigma$ expresses the measurement and model uncertainties.


%
%

To evaluate the complexity of an appliance set, it is now possible to apply the introduced disaggregation complexity for each possible combined power value.
This yields information which power values and therefore appliance state combinations are more complex than others.

\begin{figure*}
\centering
\includegraphics[width=1\columnwidth]{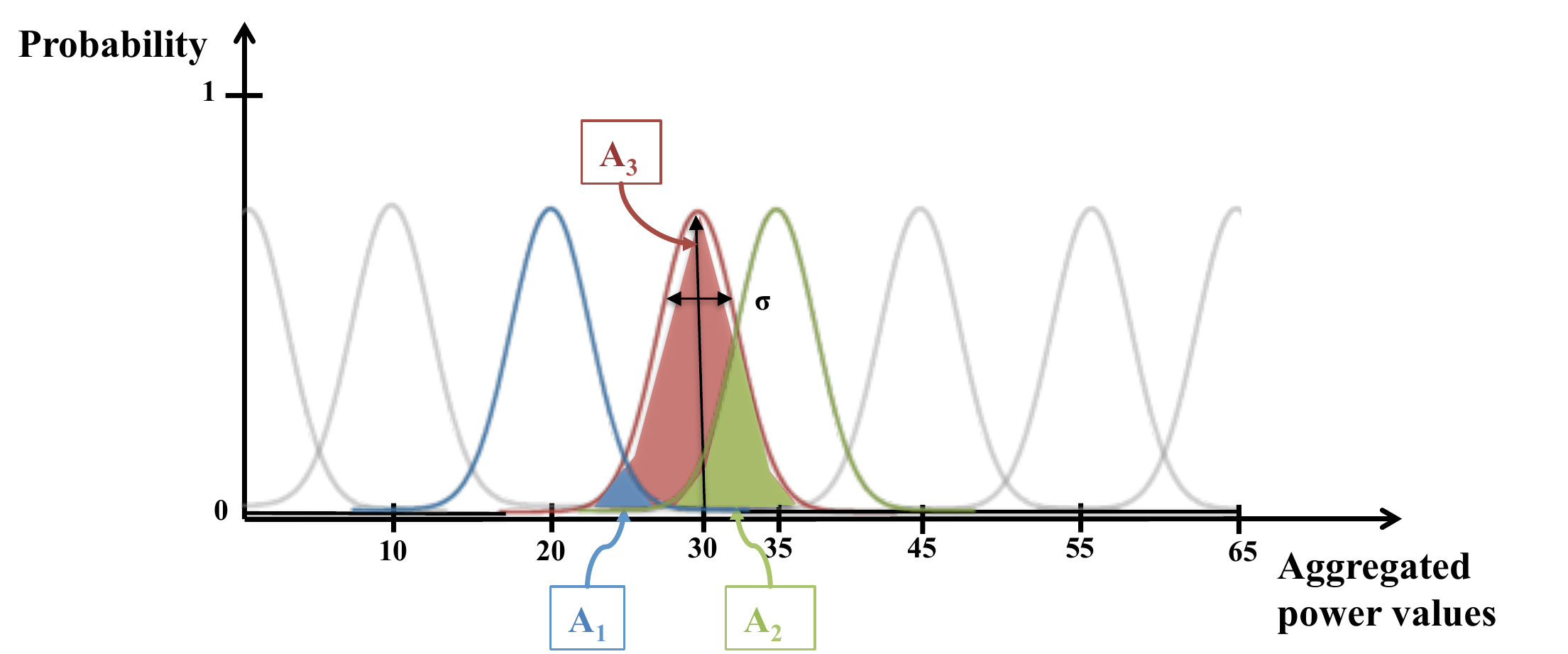}
\caption{A sketch of the different \ac{PDF}s for each power value produced by the combination of all available power demands of an appliance set. The appliance set consists of three on-off appliances with demands of $10$, $20$ and $35W$. }
\label{fig:pdfComplexNew}
\end{figure*}

Figure \ref{fig:pdfComplexNew} sketches an example how to estimate the disaggregation complexity. 
For a given set of three on-off devices with \{10,20,35\}W we estimate the complexity for the power value $P_k$ of $30$ that represents the case when device one and two are turned on.  
The set has $M=8$ possible power values in total.
%
Each power state is represented by the same normal distributed \ac{PDF}.
%
%
%
The final disaggregation complexity value is then the sum of all overlapping areas, like $A_1$, $A_2$ and $A_3$ shown in Figure \ref{fig:pdfComplexNew}.
The introduced disaggregation complexity $C$ can be interpreted as a similarity factor of power states in the appliance set.

Accordingly, a disaggregation complexity $C$ of $1$ means that at least one solution or appliance state is equal to the wanted power value. But it can also mean that two power value distributions match with similarity $0.5$.
The disaggregation complexity $C=2$ means that in the case of two appliances each of them has the same power demand.
Exceptions are the all-off power state (0W) and the maximum power demand $P_M$.
Through to bounds of the complexity computation by $[0,P_M]$ these states show a value of $C = 0.5$.
The values of $C$ depend as well on the chosen variance $\sigma$ of the \ac{PDF}.
The higher the value of $\sigma$, the higher is the probability of intersections between power values.
This means the higher $\sigma$ the higher is the appliance set complexity.
A whole appliance set is characterized by its power states complexity spectrum that shows the complexity value for each of the aggregated power state values.
The power states complexity spectrum shows at which regions confusions of states and therefore wrong appliance detections are more likely.

\subsection{Time Series Complexity of Aggregated Power Profiles}
The introduced disaggregation complexity $C$ considers the appliance set and its characteristics but does not refer to a specific aggregated power profile.
Therefore, we introduce the time series disaggregation complexity $C_{total}$ which is a weighted average of the complexities of the power values within a time series.
It considers the appliance set implicitly through the disaggregation complexity.
The usage of the different appliances is reflected by the power values in the profile.
We define the time series disaggregation complexity of an aggregated power draw as 	
%
\begin{equation}
C_{total} = \frac{1}{T} \sum_{t=1}^{T} C_{t} = \frac{1}{T} \sum_{t=1}^{T} \sum_{k=1}^M \mathrm{OVL}(f_{P_t},f_{P_k}) \quad,
\end{equation}
where $T$ represents the number of observed power samples.
This disaggregation complexity $C_{total}$ describes the averaged complexity of observed power values within all possible appliance state combinations for the whole observation time.
The same power profile can have a different total complexity with respect to the appliance set.
That reflects exactly the difficulty in load disaggregation to distinguish between similar devices.
Calculation of $C_{total}$ requires knowledge of the respective appliance set, i.e., their number of states, the power values and their distribution (or reasonable assumptions about it).

Application of the total complexity to a power profile that lacks any meta-data requires additional assumptions. Those must be mentioned to keep the complexity understandable.
The first attempt therefore would be to analyze the power profile histogram to reason about the aggregated states and distributions.
Distortion can occur if two values close to each other are considered as different states instead of including them in the same distribution. This approximately doubles the single value complexity and consequently leads to overestimation of the total profile complexity.
For instance, small power values (which are hardly distinguishable from noise) can be smoothed away by a threshold to avoid many very similar states that cause a misleading high complexity number.

\section{Evaluation Settings} \label{sec:evaluationSettings}

\subsection{Real World Dataset} \label{subsec:applianceDataset}
To test the disaggregation complexity metric on different test cases we performed our complexity study on three different datasets.
The first choice is the open {REDD} dataset \cite{kolter-kdd-2011}.
We choose three different houses from the dataset where 6 appliances were selected according to its characteristics to affect the household power demand in a significant way \cite{Carlson2013132}.
Furthermore, we used the open dataset {GREEND} \cite{Andrea2014}, which documents an appliance level measurement campaign in Austria and Italy.
As for the {REDD} dataset we have chosen 3 houses with 6 different appliance as representative for our evaluation.
The {ECO}-Dataset \cite{Beckel2014} was also used which monitored electricity consumption and occupancy in 9 Swiss houses.
3 houses with 6 different appliances were chosen.
In Table \ref{tab:applianceSetComplexity} the appliances for each house and dataset are listed.
For our evaluation we have chosen the whole observation time for the {REDD} dataset and two week for the {GREEND} and {ECO}-dataset.
This assumptions are valid through the whole paper if not mentioned in a different way.

\begin{landscape}
\begin{table*}[htpb]
\centering
\begin{tabular}{ll|p{5cm}p{5cm}p{5cm}c}
\toprule
 \textbf{Dataset} & \textbf{House} & \textbf{Appliance Type} &  \textbf{Detected Power (submetered)} & \textbf{Detected power (aggregated)} \\
\midrule
\textit{{REDD}} & \textit{1} & oven, fridge, dishwasher, microwave, stove, washer dryer  & [1680 2478], [200 420], [50 210 410 890 1115], [55 110 270 300 620 1405 1505], [260 710 1440], [2705] &  [55], [200], [250], [410], [710], [890], [1078], [1368], [1620], [17425], [2270], [2504], [2670]\\

\textit{{REDD}} & \textit{2} & kitchen outlet 1, lighting, stove, microwave, kitchen outlet 2, fridge & [130 210 770], [123], [410], [40 1718 1850], [1050], [160 420] & [90], [145], [245], [310], [410], [600], [770], [937], [1060], [1752], [1885]\\

\textit{{REDD}} & \textit{3} & fridge, dishwasher, washer dryer, microwave, bathroom gfi, kitchen outlet & [100 400], [210 525 730], [2265], [120 540 1698], [860 960 1285 1605], [40 365 900 1220 1520] & [70], [120], [205], [270], [370], [535], [730], [960], [1274], [1676], [1835], [2197], [2367], [2630] \\
\midrule
\textit{{ECO}} & \textit{1} & fridge, dryer, coffee machine, kettle, washing machine, PC & [40], [250 440 785], [50 1225], [1800], [90 180 250 365 21688], [72] & [105], [245], [335], [545], [900], [1232], [1800], [2170]\\
\textit{{ECO}} & \textit{2} & diswasher, fridge, entertainment (stereo system  and TV), Freezer, water kettle, dimmable lamp & [120 2132], [70], [55 175], [50 310], [50 1840], [80 185] & [110], [190], [280], [510], [18689], [2108]\\
\textit{{ECO}} & \textit{3} & fridge, kitchen appliances (coffee machine, bread baking machine and toaster), lamp, freezer, entertainment (stereo and TV), microwave & [100], [67 190 280 445 650 785 1065 1545], [130], [100 175 280], [120], [40 1365 1485] & [80], [135], [195], [265], [435], [668], [841], [1007], [1185], [1386], [1565] \\
\midrule
\textit{{GREEND}} & \textit{1} &   coffee machine, washing machine, fridge, dishwasher, water kettle, vacuum cleaner & [60 148 470 570 1225 1265], [70 155 210 260 423 1898], [55 140 240], [40 1900], [1790], [1220] & [110], [239], [448], [540], [1267], [18967] \\
\textit{{GREEND}} & \textit{2} & fridge, dishwasher, water kettle, washing machine, dryer, bedside light & [80], [80 1725], [850], [90 173 1910], [1580], [60] & [92], [182], [845], [1583], [1775], [1900] \\
\textit{{GREEND}} & \textit{3} & TV, washing machine, dryer, dishwasher, kitchenware, coffee machine  & [110 235 285 360], [125 245 358 1998 2100], [70 160 2358 2550], [70 2002], [120 1235], [55 125 540 882 1047 1220 1630] & [110], [295], [530], [863], [1043], [1230], [1635], [1920], [2093], [2355], [2554], [2830]   \\
\bottomrule
\end{tabular}
\caption{List of datasets ({REDD}, {ECO}-dataset, {GREEND}) with 6 chosen appliances and their appliance power states detected for submetered power draws and the aggregated power draw.}
\label{tab:applianceSetComplexity}
\end{table*}
\end{landscape}

\subsection{Identification of Appliance Power States}\label{subsec:ident}
To be able to compute the two complexity measures, the set of occurred power states is necessary.
If meta data provides this information, this data could be used, but for most cases and datasets this information is either not provided or not in the desired extent.
Accordingly, the most obvious approach would be to use expert knowledge to identify the appliance states and their power demand.
But this process is time consuming and erroneous.
Therefore, an automatic state detection algorithm is presented, based on the approach published in \cite{Egarter2015}.
It automatically detects the most common power states in any used power draw.
In detail, the approach starts with de-noising the signal from high frequency signal by median filtering.
The signal is sharpened to get sharp edges which is helpful and also necessary for edge detection.
The edge detection detects based on a predefined threshold rising and falling edges in the signal.
With the set of rising edges and the set of falling edges, edge pairs are identified.
These detected edge pairs are then the basis for frequency counting of edges done be performing an histogram of occurred edge pairs.
Based on the histogram, it is decided if an edge pair is representative for the considered observation window.
More and detailed information about the detection process can be found in \cite{Egarter2015}.

The detection approach can be applied on submetered measurement data as well as on aggregated power measurements.
For both scenarios different outputs are produced in which the submetered measurements can produce multi-state power states of appliances.
Similarities between appliances and their power states are possible.
In contrast the aggregated power measurement data is producing a set of power states without any information of appliances and their number of states.
It is only detecting different power states and not different appliances.
Considering this input case, no similarities between appliances are possible.
The algorithm tries to find a unique set of power states.
However, we want to clarify that the use of this detection approach is not necessary for the calculation of the complexity values.
The complexity values can be applied to any detection approach providing a set of appliances power states in which the appliances are described as on/off or multi-state appliances.\newline
In this work the proposed detection approach is applied to each house and dataset.
The results are listed in Table \ref{tab:applianceSetComplexity}.
The parameters of the algorithm are set as in \cite{Egarter2015} as for example the power threshold for a valid power edge was $25W$.

\subsection{Load Disaggregation Algorithms}\label{subsec:loadDisaggregationAlgorithm}
The proposed complexity values should describe the complexity of aggregated power loads.
To get an idea how meaningful the proposed complexity approaches are, the results should be compared to the results of an appropriate and suitable load disaggregation approach.
This comparison should give a quantitative feedback if the complexity value is meaningful according to the used load disaggregation approach.
We claim that the load disaggregation approach needs to have the same inputs as described in Section \ref{sec:Complexity} to be able to provide meaningful results.\newline
Thus in the following, we review state-of-the-art load disaggregation algorithms which can be divided into supervised and unsupervised algorithms.
Supervised \ac{NILM} approaches are based on labelled data on which pattern recognition and optimization algorithms are applied \cite{Egarter2013, Suzuki2008, Zeifman2012, Srinivasan2006, Lin2010,Hassan2014,Lin2014,wang2012}.
Labelled data and the corresponding need of a training phase is the main disadvantage of supervised algorithms, because of increased development costs and efforts.
In contrast to supervised algorithms, unsupervised algorithms do not need labelled data as well as no training phases.
Recent representatives for unsupervised algorithms are based on k-means clustering \cite{goncalves_unsupervised_2011}, on fractional hidden Markov models (FHMM) and its variants \cite{zico2012,Zaidi2010,Zia2011,Kim2011,Zoha2013}.
In this work, the \ac{PF} approach to perform load disaggregation is used \cite{Egarter2014A}.
The \ac{PF} is beneficial for \ac{NILM} because of its characteristics to handle non-linear problems within large state spaces that are suffering from non-Gaussian noise.
Basically, the \ac{PF} estimates the power states of appliances based on the approximated posterior density of power states estimated by a random set of particles.
The appliance state space is composed by multiple independent \ac{HMM} representing on the one hand appliances such as on/off or multi-state appliances and on the other hand the aggregated power draw for a household power demand.
The composition of \ac{HMM}s leads to the use of a \ac{FHMM} which has the advantage to decrease the number of states compared to using one large standard \ac{HMM}.
As a final step, the work in \cite{Egarter2014A} uses a decision maker based on thresholding to decide which appliance is working at which operating state with knowledge of the appliance \ac{HMM}.
The appliance model includes on the one hand the approximated power demand and on the other hand the general structure, such as how many states a device has.
The \ac{PF} is able to work with unknown and inaccurate transition matrix settings.
The \ac{PF} based approach is suitable for our evaluation due to the fact that the algorithm can handle a set of appliances modelled as on/off or multi-state appliances and is performing load disaggregation based on a set of power states and the aggregated power draw.
For the evaluation the \ac{PF} is parametrized as in \cite{Egarter2014A} in which the number of used particles, as most important parameter, is set to $1000$ particles.

\section{Case Study} \label{sec:caseStudy}

\subsection{Appliance Set Complexity for Different Datasets and Different Sets of Power States }
\begin{landscape}
\begin{figure*}[htpb]
\centering
	\includegraphics[width=1\columnwidth]{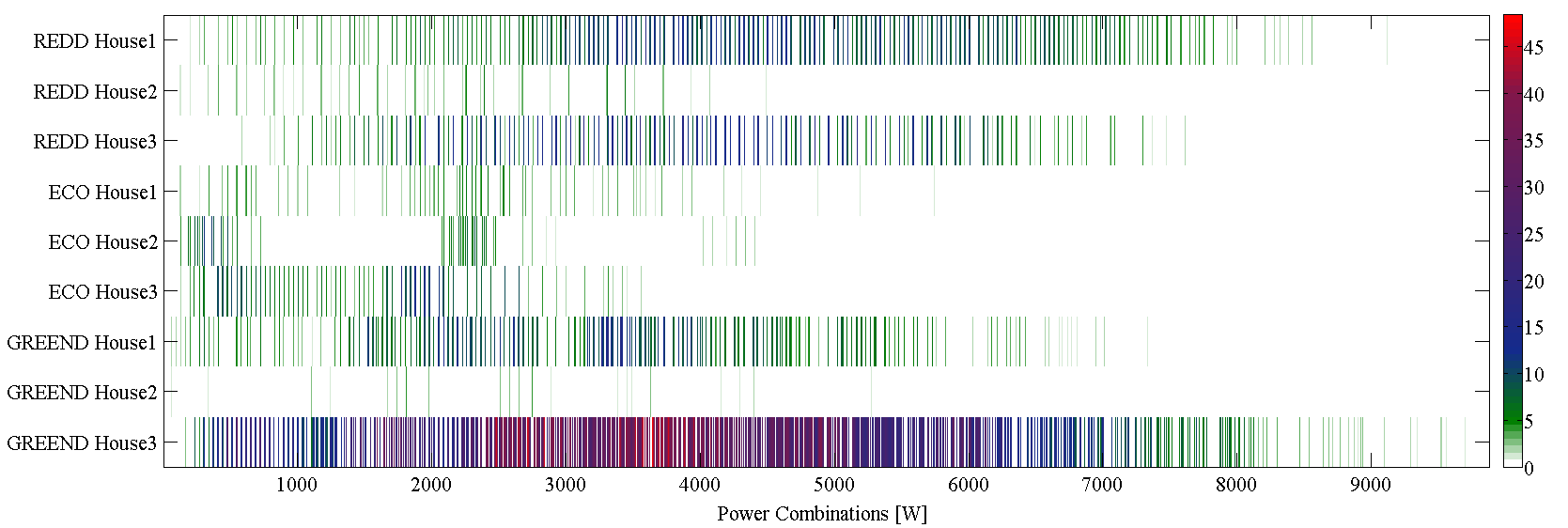}
	\caption{Colormap of the appliance set complexity for 3 houses of the {REDD}, {ECO}, {GREEND} dataset over the possible power combination of all houses and datasets. 0 means that power value is not possible for the used house and dataset. Therefore no complexity value is applicable to this power combination}
		\label{fig:AC_submetered}
\end{figure*}
\end{landscape}

As described in the previous sections, the appliance set complexity is aiming to describe the complexity of the used appliance set without considering the appliance usage over time.
Therefore, the most relevant parameter are the used power values for each appliance power state.
These power states are identified for each appliance using the algorithm presented in Section \ref{subsec:ident}.
The algorithm creates power states from measurement data in which we used aggregated measurement data and submetered power readings.
In the case of aggregated power readings the algorithm has no information about the number of appliances.
Thus, it can distinguish between different power states but not between different appliances.
Appliance with similar power consumption are handled as single on/off devices and multi-state behavior of appliances is not assigned to appliances.
For the evaluation the aggregated power consumption is created by superimposing the appliance power draws from appliances chosen in Table \ref{tab:applianceSetComplexity}.
\begin{table}[h!]
\centering
\begin{tabular}{ll|cccc}
\toprule	
\multirow{3}{1cm}{\textbf{Dataset}}& \multirow{3}{0.8cm}{\textbf{House}} &\multicolumn{2}{c}{\textbf{submetered}}&\multicolumn{2}{c}{\textbf{aggregated}}	\\
\cmidrule(r){3-4} \cmidrule(r){5-6}
 && \textit{max} & \textit{mean}  &  \textit{max} & \textit{mean} \\
\midrule
  \textit{{REDD}} & \textit{1} & 16.91 &  7.88  & 2.28 & 1.48  \\
  \textit{{REDD}} & \textit{2} & 6.170 & 2.62  & 2.32 & 1.33 \\
  \textit{{REDD}} & \textit{3} & 21.39 & 8.69  &1.98 & 1.32   \\
\midrule
  \textit{{ECO}} & \textit{1} & 6.65 & 2.88  & 2.67 & 1.36   \\
  \textit{{ECO}} & \textit{2} & 12.06 & 4.75  & 1.44 & 1.04   \\
  \textit{{ECO}} & \textit{3} & 16.62 & 6.53  & 1.59 & 1.15  \\
\midrule
  \textit{{GREEND}} & \textit{1}  & 18.20 & 7.17  & 2.01 & 1.19 \\
  \textit{{GREEND}} & \textit{2} & 4.46 & 2.18   & 1.36 & 1.07  \\
  \textit{{GREEND}} & \textit{3} & 48.36 & 24.43  & 1.87 & 1.18  \\
\bottomrule
\end{tabular}
\caption{List of mean and maximum of the appliance set complexity for each house and dataset}
\label{tab:Complexity}
\end{table}
In contrast, we used submetered power readings to create multi-state appliances with appliance having similar power states
With this consumption data of an appliance it is possible to identify appliance specific power states which can be combined to a multi-state appliance.
We performed the power state detection for each device listed in Table \ref{tab:applianceSetComplexity}.
The created set of appliances consists of on/off and of multi-state appliances in which the power states between appliances can be the same or differ only for some Watts or be completely different.
The created appliances for the aggregated consumption data and for the submetered power readings are presented for each used dataset in Table \ref{tab:applianceSetComplexity}.\newline
In this case study the appliance set complexity is tested on the appliance set based on aggregated power readings and on submetered power readings.
As input for the complexity computation a vector of all possible power state combinations of the appliance set is used.
The results are presented in Table \ref{tab:Complexity} using the mean and the maximum value of the appliance complexity.
The complexity values for submetered data are higher and therefore more complex than for the aggregated power readings.
As reason we claim that similarities between appliances are getting lost in the case of aggregated loads due to the inability to distinguish between appliances.
With aggregated power readings it is only possible to distinguish between different power states.
This also leads to the fact that the problem complexity for the same house of a dataset differs between appliance sets created by the aggregated or the submetered power data.
This strengthens the need of a complexity measure due to different preprocessing stages of power data.
However, appliances produced by submetered data are affected by power state similarities and have therefore a higher appliance set complexity.
We also provide Figure\footnote{For readability please consider coloured prints} \ref{fig:AC_submetered} presenting the appliance set complexity for each dataset over all possible power state combinations and is based on the appliance states produced by the submetered power readings.
The plot shows for each possible power state combination the appliance set complexity.
The color white means that the appliance set complexity is zero because this power value is not producible by a combination of saved power states for a certain dataset and house.
The appliance set complexity starts from green (low complexity), blue (medium complexity) and ends at red (high complexity).
The colors are normalized according to the dataset with the maximum occurred appliance set complexity.
Figure \ref{fig:AC_submetered} shows which dataset and house is more complex according to the used power states presented in Table \ref{tab:applianceSetComplexity}.
For example, house 2 of the {GREEND} dataset has a very low appliance set complexity while house 3 of the {GREEND} dataset has a very high and tight appliance set complexity.

\subsection{Time Series Complexity for Different Datasets and Different Sets of Power States}
\begin{landscape}
\begin{figure*}[htpb]
\centering
	\includegraphics[width=1\columnwidth]{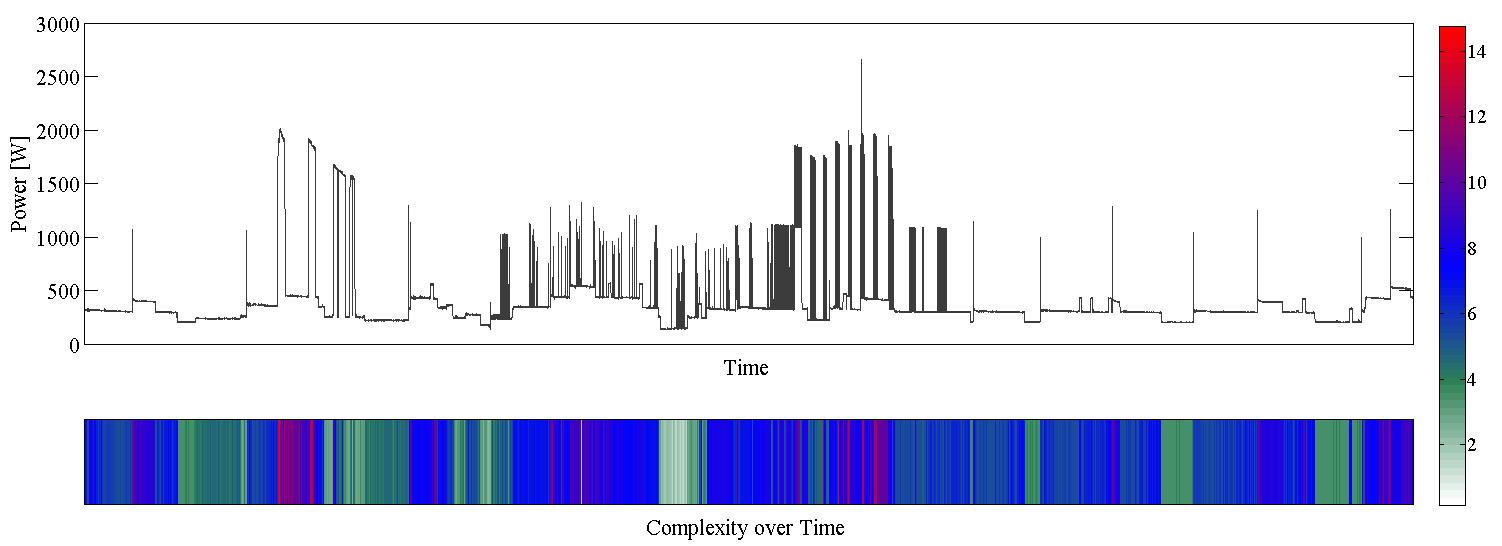}
	\caption{Time snippet of the power readings for house 3 of the {ECO} dataset with a colormap of the complexity for each measurement sample. White to green means low complexity, blue means medium complexity and red means high complexity. }
		\label{fig:TC_submetered}
\end{figure*}
\end{landscape}

The appliance set complexity gives feedback about the complexity of the used appliances by comparing their power states and appliance structure.
For the load disaggregation problem another important factor is the influence of the appliance usage over time.
This considers how and when appliances are operated which could be for example user driven (e.g., coffee machine, TV) or periodically activated (e.g., fridge).
The proposed time series complexity considers this circumstances in its computation.
For the evaluation of this complexity measure the time series of all houses and datasets for an observation window of half day are considered.
The input for the complexity computation are the measurement samples which are combinations of possible power states affected by noise.
In contrast, the appliance set complexity considers power state combination without noise as input for the complexity computation.
As appliance set the appliances based on aggregated and submetered power data are used.
In Table \ref{tab:TC_Complexity} the mean and the maximum of the time series complexity for all houses and datasets are presented.
Moreover, a time snippet of a time series of house 3 of the {ECO} dataset with corresponding complexity values for each measurement sample are presented in Figure\footnote{For readability please consider coloured prints} \ref{fig:TC_submetered}.
The colors white and green means low complexity, blue means medium complexity and red means high complexity.
The colouring is normalized to maximum occurred complexity value for the considered observation time and measurement samples.
Comparing the colormap with the time series shows that overlapping behavior results in an increased and high complexity value while high power values do not necessarily results in a high complexity.

\begin{table}[h!]
\centering
\begin{tabular}{ll|cccc}
\toprule	
\multirow{3}{1cm}{\textbf{Dataset}}& \multirow{3}{0.8cm}{\textbf{House}} &\multicolumn{2}{c}{\textbf{submetered}}&\multicolumn{2}{c}{\textbf{aggregated}}	\\
\cmidrule(r){3-4} \cmidrule(r){5-6}
 && \textit{max} & \textit{mean}  &  \textit{max} & \textit{mean} \\
\midrule
\textit{{REDD}} & \textit{1} & 13.79 & 1.04   & 1.62& 0.50   \\
\textit{{REDD}} & \textit{2}  & 5.39 & 0.54    & 2.32& 0.11  \\
\textit{{REDD}} & \textit{3} & 17.54 & 1.07   & 1.98& 0.35   \\
\midrule
\textit{{ECO}} & \textit{1}  & 3.71 & 0.95   & 2.62& 0.15   \\
\textit{{ECO}} & \textit{2}  & 11.99& 2.86  & 1.11&  0.19   \\
\textit{{ECO}} & \textit{3}  & 14.77 & 4.91 & 1.57&  0.41   \\
\midrule
\textit{{GREEND}} & \textit{1} & 7.77 & 0.89  & 1.06& 0.12   \\
\textit{{GREEND}} & \textit{2}  & 4.305 & 0.91  & 1.35& 0.50  \\
\textit{{GREEND}} & \textit{3}  & 45.01& 3.67   & 1.81& 0.04 \\
\bottomrule
\end{tabular}
\caption{List of mean and maximum of the time series complexity for each house and dataset}
\label{tab:TC_Complexity}
\end{table}

\subsection{Load Disaggregation of Complexity Marked Power Readings}
In this case study the results of the complexity measures are compared with the results of an \ac{NILM} approach on the same power data.
The aim is not to evaluate the used disaggregation approach.
This evaluation should give a feedback about the suitability and meaningfulness of the proposed complexity measures.
As described in Section \ref{sec:evaluationSettings} we used the load disaggregation algorithm from \cite{Egarter2014A} which is able to handle on/off and multi-state appliances.
We used the appliance set and models identified by the submetered measurements from Table \ref{tab:newApplianceSetComplexity}.
\begin{table}[h!]
\centering
\begin{tabular}{ll|p{5.5cm}}
\toprule	
\textbf{Dataset} & \textbf{House} & Appliance States \\
\midrule
\textit{{REDD}} & 1  & [1690 2455], [190] [210 410 880 1110], [60 1533], [260 710 1440] [2712]\\
\textit{{REDD}} & 2  & [770], [145], [410], [1875], [1050], [160] \\
\textit{{REDD}} & 3  & [120], [210] [2255], [130 1740], [960 1290 1610], [360 900] \\
\midrule
\textit{{ECO}} & 1  & [40], [780], [50 1205], [1795], [80], [90] \\
\textit{{ECO}} & 2  &  [120 2060 2170], [70], [55 178], [50], [1845], [160]\\
\textit{{ECO}} & 3  & [100], [55 1085 1520], [130], [100], [120], [1330 1567]\\
\midrule
\textit{{GREEND}} & 1 & [50 1270], [55 1840], [50 140], [40 1900], [1790], [1220]\\
\textit{{GREEND}} & 2 & [80], [80 1730], [850], [90 160 1910], [1580], [60]\\
\textit{{GREEND}} & 3 & [60], [72 2020], [160 2415], [70], [1230], [1030] \\

\bottomrule
\end{tabular}
\caption{Appliance set used by the load disaggregation approach. The appliances differ to the ones in Table \ref{tab:applianceSetComplexity} due to the fact that we considered for this case study only the most common power states per device}
\label{tab:newApplianceSetComplexity}
\end{table}

We assume the availability of ground truth data for the evaluation as reason to use the submetered data and not the aggregated power readings.
The appliance set detected in Table \ref{tab:newApplianceSetComplexity} compared to the listed ones in Table \ref{tab:applianceSetComplexity} are different because the appliance state identification algorithm from Section \ref{sec:evaluationSettings} was considering only the most common appliance power states.
We defined power states as most common appliance power states if a detected power state occurred as often as $15\%$ of the maximum occurred power state.
We used power readings of a whole day to calculate the time-series complexity.
The load disaggregation algorithm is evaluated according to the real and estimated energy per $kWh$ on appliance level and to the aggregated power readings.
The results for each house and dataset for all used appliances are shown in Table \ref{tab:NILM}.
\begin{landscape}
\begin{table*}
\centering
\begin{tabular}{lp{0.7cm}|cccccc|ccc}
\toprule	
\textbf{Dataset} & \textbf{House} & App. 1 & App. 2 & App. 3 & App. 4 & App. 5 & App. 6 & \textbf{Total} & \textbf{AC} & \textbf{TC} \\
 & & \textit{real/est.} & \textit{real/est.} & \textit{real/est.} & \textit{real/est.} & \textit{real/est.} & \textit{real/est.} & \textit{real/est.} & \textit{mean/max} & \textit{mean/max} \\
\midrule
\textit{{REDD}} & 1 & 0.13/0.22  &  1.27/0.98  &  0.31/0.43 &   0.53/0.21 &   0.003/0.32  &  0.0/0.06 & 2.23/2.21 & 2.97/9.06 & 0.41/4.64\\
\textit{{REDD}} & 2 & 0.19/0.13  &   0.82/0.99 &   0.05/0.28 &    0.29/0.05&    0.24/0.20 &   1.67/1.44 & 3.26/3.01 & 2.01/4.69 & 0.23/1.27\\
\textit{{REDD}} & 3 &1.08/0.94   & 0.16/0.25  &   0.70/0.78 &   0.20/0.29 &   0.69/0.87 &   0.33/0.34 & 3.17/3.46 & 1.69/3.78 &  0.40/4.09\\
\midrule
\textit{{ECO}} & 1 & 0.54/0.35 &   0.001/0.04 &   0.23/0.26 &   0.0002/0.02 &   0.002/0.34 &   0.49/0.26 & 1.27/1.27 & 1.469/2.69 & 0.84/2.59\\
\textit{{ECO}} & 2 & 0.0/0.05 &   0.53/0.61 &   0.86/0.067 &   0.71/0.54 &   0.30/0.31 &   0.01/0.82  & 2.39/2.40 & 2.72/5.83 & 0.758/3.038\\
\textit{{ECO}} & 3 & 0.66/1.18 &   0.48/0.32 &   0.073/1.55 &   4.18/1.26 &   0.54/1.46 &   0.42/0.48 & 6.30/6.25 & 2.34/6.45 & 0.54/2.66\\
\midrule
\textit{{GREEND}} & 1 &0.11/0.29 &   0.0/0.10  &  1.20/0.32  &  0.01/0.41 &   0.0/0.03 &   0.0/0.081& 1.32/1.24 & 2.57/6.04 & 1.08/5.15\\
\textit{{GREEND}} & 2 & 0.55/0.43  &  0.81/0.04   &      0.0/0.03 &   0.0/0.04  &  0.19/0.82 &   0.0/0.196 & 1.56/1.55 & 1.07/1.27 & 1.002/3.023 \\
\textit{{GREEND}} & 3 & 2.59/0.49  &  0.93/0.94   & 1.94/1.60 &   0.65/0.58 &   0.08/1.50 &   0.19/1.40 & 6.37/6.48 & 1.73/4.01 & 0.42/2.15\\
\bottomrule
\end{tabular}
\caption{List of the load disaggregation result (real and estimated) on appliance level and in total for all houses and datasets. For comparison also the appliance set complexity (AC) and time-series complexity (TC) are shown.}
\label{tab:NILM}
\end{table*}
\end{landscape}

Less complex time series like in {REDD} house 2 are easier to disaggregate than more complex time series as in {ECO} house 2.
A lower complexity is in general easier to disaggregate as a more complex time series.
Similar power states as for example in house 1 and 2 in the {ECO} dataset are highly affecting the load disaggregation result.
In the case of similar power states the algorithm is not able to distinguish between appliances with similar power states which is supporting the need of a common complexity measure for load disaggregation.
By using a different power state identification setting also the appliance set complexity compared to the previous case studies is different.
This also strengthens our assumption to have a complexity measure handling the set of appliance power states independent from the used load disaggregation algorithm.

\section{Discussion}\label{sec:discussion}
In the previous section different case studies were presented to evaluated usefulness of the proposed complexity measures.
For example in the case study for the appliance set complexity the complexity is highly dependent on the used appliance set.
The number of devices several states and similar states between appliances are affecting the load disaggregation complexity strongly.
The complexity is higher for more complex appliance sets and is not dependent on the used house or dataset.
Thus, we claim that the preprocessing stage has an important effect on the problem complexity and accordingly also on the result of the used load disaggregation process.
This fact is also valid for the time-series complexity.
By using an appliance dataset with complex appliances and structures also the time-series complexity is affected strongly for the same house of a dataset.

The time series complexity is highly affected by the appliance usage.
For the evaluation of the case study an observation window of a day was used.
We claim that even complex appliance sets as the house 3 of the {GREEND} dataset can have a low time series complexity due their appliance usage over time.
Thus, the appliance set complexity and the time series complexity do not correlate between each other.
For example a high appliance set complexity can lead to a low or a high time series complexity.
We also show that the proposed complexity measures can classify the complexity of a load disaggregation problem but does not correlate to the used load disaggregation approach.
The result of the load disaggregation approach cannot be estimated by our proposed approach in which the problem with its preprocessing stages is described and made comparable.
However the case studies showed that the proposed complexity measures fulfil the requirements identified in Section \ref{sec:Complexity}.
It uses a given appliances dataset to compute the appliance set complexity as well as the time series complexity.
The complexity measures can be interpreted as a similarity measure between a set of possible states (appliance set complexity) and a similarity measure in which the set of possible states is compared with noisy observed data (time series complexity). 
The same principle for measuring complexity is applicable for other attributes.

\section{Conclusion} \label{sec:conclusion}
This paper defined two complexity measures for the problem of load disaggregation which deals with the task to break down the aggregated power draw of appliance to the appliance components.
Appliance characteristics with smart algorithms are used to solve this task.
One important aspect is the distinction between the disaggregation approach itself and the problem of aggregated power profiles.
Beside clear performance measures for \ac{NILM} algorithms it needs a clear definition to specify the hardness or complexity of a specific case.
This makes a fair comparison of different \ac{NILM} approaches according to the used load disaggregation problem possible.
To overcome the lack to compare load disaggregation problems we introduced two novel complexity measures to assess the complexity of a load disaggregation problem based on the used appliance sets.
With the proposed complexity measures the used appliance sets and the aggregated power readings are evaluated for their complexity.
To evaluate how the disaggregation complexity measures are reflecting load disaggregation problems in reality, we performed the complexity calculation and load disaggregation with a state-of-the-art \ac{NILM} approach on different datasets and time-series.
Our evaluations show that our disaggregation  complexity measure is able to assess the hardness of an appliance dataset as well as a specific time series for a NILM algorithm.
We want to emphasize that the presented complexities are relative and not absolute measures for the problem complexity.
Thus, knowing the disaggregation complexity is not sufficient to determine the performance of the load disaggregator as the performance to disaggregate loads depends on.
The presented measure gives meaningful results for load disaggregation problems with one feature such as the active power representing each power state of an appliance in which future work will deal with several features as active and reactive power.

\bibliographystyle{IEEEtran}
\bibliography{NIALM}

\begin{thebibliography}{10}
\providecommand{\url}[1]{#1}
\csname url@samestyle\endcsname
\providecommand{\newblock}{\relax}
\providecommand{\bibinfo}[2]{#2}
\providecommand{\BIBentrySTDinterwordspacing}{\spaceskip=0pt\relax}
\providecommand{\BIBentryALTinterwordstretchfactor}{4}
\providecommand{\BIBentryALTinterwordspacing}{\spaceskip=\fontdimen2\font plus
\BIBentryALTinterwordstretchfactor\fontdimen3\font minus
  \fontdimen4\font\relax}
\providecommand{\BIBforeignlanguage}[2]{{%
\expandafter\ifx\csname l@#1\endcsname\relax
\typeout{** WARNING: IEEEtran.bst: No hyphenation pattern has been}%
\typeout{** loaded for the language `#1'. Using the pattern for}%
\typeout{** the default language instead.}%
\else
\language=\csname l@#1\endcsname
\fi
#2}}
\providecommand{\BIBdecl}{\relax}
\BIBdecl

\bibitem{Hart1992}
G.~Hart, ``{Nonintrusive appliance load monitoring},'' \emph{Proceedings of the
  {IEEE}}, vol.~80, no.~12, pp. 1870--1891, 1992.

\bibitem{Pincus1991}
S.~M. Pincus, ``Approximate entropy as a measure of system complexity.''
  \emph{Proceedings of the National Academy of Sciences}, vol.~88, no.~6, pp.
  2297--2301, 1991.

\bibitem{Richman2000}
J.~S. Richman and J.~R. Moorman, ``Physiological time-series analysis using
  approximate entropy and sample entropy,'' \emph{American Journal of
  Physiology - Heart and Circulatory Physiology}, vol. 278, no.~6, pp.
  H2039--H2049, 2000.

\bibitem{Rosso2003}
A.~P. Rosso~OA, MT~Martin, ``Brain electrical activity analysis using wavelet
  based informational tools,'' \emph{Physica A}, 2002.

\bibitem{Monge2014}
R.~E. Monge and J.~L. Crespo, ``Comparison of complexity measures for dna
  sequence analysis,'' in \emph{Bio-inspired Intelligence (IWOBI), 2014
  International Work Conference on}, July 2014, pp. 71--75.

\bibitem{Costa2005}
M.~Costa, A.~Goldberger, and C.-K. Peng, ``Multiscale entropy analysis of
  biological signals,'' \emph{Phys. Rev. E}, vol.~71, p. 021906, Feb 2005.

\bibitem{Rezek1998}
I.~Rezek and S.~Roberts, ``Stochastic complexity measures for physiological
  signal analysis,'' \emph{Biomedical Engineering, IEEE Transactions on},
  vol.~45, no.~9, pp. 1186--1191, Sept 1998.

\bibitem{Jäntti2004}
V.~Jäntti, S.~Alahuhta, J.~Barnard, and J.~W. Sleigh, ``Spectral
  entropy—what has it to do with anaesthesia, and the eeg?'' \emph{British
  Journal of Anaesthesia}, vol.~93, no.~1, pp. 150--152, 2004.

\bibitem{Gao2012}
J.~Gao, J.~Hu, and W.-w. Tung, ``\BIBforeignlanguage{English}{Entropy measures
  for biological signal analyses},''
  \emph{\BIBforeignlanguage{English}{Nonlinear Dynamics}}, vol.~68, no.~3, pp.
  431--444, 2012.

\bibitem{Dong2014}
R.~Dong, L.~Ratliff, H.~Ohlsson, and S.~S. Sastry, ``Fundamental limits of
  nonintrusive load monitoring,'' in \emph{Proceedings of the 3rd International
  Conference on High Confidence Networked Systems}, ser. HiCoNS '14.\hskip 1em
  plus 0.5em minus 0.4em\relax New York, NY, USA: ACM, 2014, pp. 11--18.

\bibitem{prokop14}
D.~Egarter, C.~Prokop, and W.~Elmenreich, ``Load hiding of household's power
  demand,'' in \emph{Proc. IEEE International Conference on Smart Grid
  Communications (SmartGridComm'14)}, Venice, Italy, 2014.

\bibitem{Cover2005}
T.~M. Cover and J.~A. Thomas, \emph{{Elements of Information Theory Wiley
  Series in Telecommunications and Signal Processing}}.

\bibitem{Gruenwald2004}
\BIBentryALTinterwordspacing
P.~Gr{\"{u}}nwald and P.~M.~B. Vit{\'{a}}nyi, ``Shannon information and
  kolmogorov complexity,'' \emph{CoRR}, vol. cs.IT/0410002, 2004. [Online].
  Available: \url{http://arxiv.org/abs/cs.IT/0410002}
\BIBentrySTDinterwordspacing

\bibitem{Inman1989}
H.~F. Inman and E.~L. Bradley, ``{The overlapping coefficient as a measure of
  agreement between probability distributions and point estimation of the
  overlap of two normal densities},'' \emph{Communications in Statistics -
  Theory and Methods}, vol.~18, no.~10, pp. 3851--3874, Jan. 1989.

\bibitem{kolter-kdd-2011}
J.~Z. Kolter and M.~J. Johnson, ``{REDD: A Public Data Set for Energy
  Disaggregation Research},'' in \emph{Proceddings of the {SustKDD} Workshop on
  Data Mining Applications in Sustainability}, 2011.

\bibitem{Carlson2013132}
D.~R. Carlson, H.~S. Matthews, and M.~Berg�s, ``One size does not fit all:
  Averaged data on household electricity is inadequate for residential energy
  policy and decisions,'' \emph{Energy and Buildings}, vol.~64, no.~0, pp. 132
  -- 144, 2013.

\bibitem{Andrea2014}
A.~Monacchi, D.~Egarter, W.~Elmenreich, S.~D’Alessandro, and A.~M. Tonello,
  ``{GREEND}: an energy consumption dataset of households in {I}taly and
  {A}ustria,'' in \emph{Proc. IEEE International Conference on Smart Grid
  Communications (SmartGridComm'14)}, Venice, Italy, 2014.

\bibitem{Beckel2014}
C.~Beckel, W.~Kleiminger, R.~Cicchetti, T.~Staake, and S.~Santini, ``The eco
  data set and the performance of non-intrusive load monitoring algorithms,''
  in \emph{Proceedings of the 1st ACM Conference on Embedded Systems for
  Energy-Efficient Buildings}, ser. BuildSys '14.\hskip 1em plus 0.5em minus
  0.4em\relax New York, NY, USA: ACM, 2014, pp. 80--89.

\bibitem{Egarter2015}
D.~Egarter and W.~Elmenreich, ``Autonomous load disaggregation approach based
  on active power measurements,'' in \emph{Proc. IEEE Workshop on Pervasive
  Energy Services (PerEnergy)}, St. Louis, Missouri, USA, 2015.

\bibitem{Egarter2013}
D.~Egarter, A.~Sobe, and W.~Elmenreich, ``Evolving non-intrusive load
  monitoring,'' in \emph{Proceedings of the Applications of Evolutionary
  Computation Conference}, 2013.

\bibitem{Suzuki2008}
K.~Suzuki, S.~Inagaki, T.~Suzuki, H.~Nakamura, and K.~Ito, ``Nonintrusive
  appliance load monitoring based on integer programming,'' in
  \emph{Proceedings of International conference on Instrumentation, Control,
  Information Technology and System Integration {(SICE)}}, 2008.

\bibitem{Zeifman2012}
M.~Zeifman, ``Disaggregation of home energy display data using probabilistic
  approach,'' \emph{{IEEE} Trans. Consum. Electron.}, vol.~58, no.~1, pp.
  23--31, 2012.

\bibitem{Srinivasan2006}
D.~Srinivasan, W.~S. Ng, and A.~Liew, ``Neural-network-based signature
  recognition for harmonic source identification,'' \emph{{IEEE} Trans. Power
  Del.}, vol.~21, no.~1, pp. 398--405, 2006.

\bibitem{Lin2010}
G.~Lin, S.~Lee, J.~Hsu, and W.~Jih, ``Applying power meters for appliance
  recognition on the electric panel,'' in \emph{Proceedings of {IEEE}
  Conference on Industrial Electronics and Applications ({ICIEA})}, 2010.

\bibitem{Hassan2014}
T.~Hassan, F.~Javed, and N.~Arshad, ``An empirical investigation of v-i
  trajectory based load signatures for non-intrusive load monitoring,''
  \emph{Smart Grid, IEEE Transactions on}, vol.~5, no.~2, pp. 870--878, March
  2014.

\bibitem{Lin2014}
Y.-H. Lin and M.-S. Tsai, ``Non-intrusive load monitoring by novel neuro-fuzzy
  classification considering uncertainties,'' \emph{Smart Grid, IEEE
  Transactions on}, vol.~5, no.~5, pp. 2376--2384, Sept 2014.

\bibitem{wang2012}
Z.~Wang and G.~Zheng, ``Residential appliances identification and monitoring by
  a nonintrusive method,'' \emph{Smart Grid, IEEE Transactions on}, vol.~3,
  no.~1, pp. 80--92, March 2012.

\bibitem{goncalves_unsupervised_2011}
H.~Goncalves, A.~Ocneanu, and M.~Berges, ``Unsupervised disaggregation of
  appliances using aggregated consumption data,'' in \emph{Proceedings of {KDD}
  Workshop on Data Mining Applications in Sustainability ({SustKDD})}, 2011.

\bibitem{zico2012}
Z.~Kolter and T.~Jaakkola, ``Approximate inference in additive factorial {HMMs}
  with application to energy disaggregation,'' in \emph{Proceedings of the
  International Conference on Artifical Intelligence and Statistics}, 2012.

\bibitem{Zaidi2010}
A.~A. Zaidi, F.~Kupzog, T.~Zia, and P.~Palensky, ``Load recognition for
  automated demand response in microgrids,'' in \emph{Proceedings of the 36th
  {IEEE} Conference on Industrial Electronics {IECON}}, 2010.

\bibitem{Zia2011}
T.~Zia, D.~Bruckner, and A.~Zaidi, ``A hidden markov model based procedure for
  identifying household electric loads,'' in \emph{Proceedings of Annual
  Conference on {IEEE} Industrial Electronics Society ({IECON})}, 2011.

\bibitem{Kim2011}
H.~Kim, M.~Marwah, M.~F. Arlitt, G.~Lyon, and J.~Han, ``{Unsupervised
  Disaggregation of Low Frequency Power Measurements},'' in \emph{Proceedings
  of the 11th {SIAM} International Conference on Data Mining}, 2011.

\bibitem{Zoha2013}
A.~Zoha, A.~Gluhak, M.~Nati, and M.~Imran, ``Low-power appliance monitoring
  using factorial hidden markov models,'' in \emph{Proceedings of {IEEE}
  Conference on Intelligent Sensors, Sensor Networks and Information
  Processing}, 2013.

\bibitem{Egarter2014A}
D.~Egarter, V.~P. Bhuvana, and W.~Elmenreich, ``{PALDi}: Online load
  disaggregation via particle filtering,'' \emph{{IEEE} Transactions on
  Instrumentation and Measurement}, vol.~64, no.~2, pp. 467--477, Feb 2015.

\end{thebibliography}
\end{document}